\documentstyle[preprint,eqsecnum,aps]{revtex}
\def\ddarrow{{\mkern+00mu\raise1.05ex\hbox{$\lfloor$}}\!\!\!
\rightarrow }
\def\pslash{{ p \mkern-08mu\raise-.1ex\hbox{/} }} 
\def\nslash{{ n \mkern-08mu\raise-.1ex\hbox{/} }} 
\tightenlines
\begin{document}
\draft
\title{ Spin Information from Vector-Meson Decay in Photoproduction }
\author{W.M. Kloet}
\address{ Department of Physics \& Astronomy, Rutgers University,
 Piscataway, New Jersey  08855-0849}
\author{Wen-Tai Chiang and Frank Tabakin }
\address{ Department of Physics \& Astronomy, University of Pittsburgh,
 Pittsburgh, Pennsylvania  15260} 
\date{\today}
\maketitle
\
\begin{abstract}
 For the photoproduction of vector mesons, 
 all single and double
spin observables involving vector meson
two-body decays are defined consistently in the 
$\gamma N$ center of mass.  These definitions yield  
a procedure for extracting 
physically meaningful single and double
spin observables that are subject to known
rules concerning their angle and energy evolution.
As part of this analysis, we show that measuring
 the two-meson decay of a photoproduced $\rho$ or $\phi$
  does not determine the vector meson's vector
polarization, 
but only its tensor polarization.  The vector meson decay into lepton pairs
is also insensitive to the vector meson's vector polarization, unless one
measures the spin of one of the leptons.  
Similar results are found for all double spin observables which 
involve observation of vector meson decay. To access the vector meson's vector 
polarization, one therefore needs to either measure the spin of the decay
leptons, make an analysis of the background interference effects or relate the 
vector meson's vector polarization to other accessible spin observables.
\end{abstract}
\pacs{24.70.+s, 25.20Lj, 13.60Le, 13.88.+e}
\widetext
\section{ Introduction}
\label{sec:introduction}
There is increased interest in measuring the
photo- and electro-production of vector mesons
from nucleon targets.  Precision experiments,
including coincidence measurements of the vector
meson's decay products, have been made possible by 
new high-flux, continuous beams of polarized  
electrons and photons with polarized targets, along with large angle 
spectrometers and recoil polarimeters. In this flowering of interest
in vector mesons and spin physics,
most experimental proposals follow the earlier analysis
 in which the vector meson's
decay is described in its rest frame. The production amplitude
for 
$\vec{\gamma} + \vec{N} \rightarrow \vec{V} + \vec{N}'\, ,$
i.e. the photoproduction of  vector mesons,~\footnote{The vector meson is 
designated by $\vec{V},$
which denotes either a $\rho$ or a $\phi$ meson, both of which
have two meson decay modes.  The $\rho$ decays to two pions, the
$\phi$ to two kaons. } is used to
produce a density matrix which describes the spin  
state of the produced vector meson. The
rest frame of the vector meson differs from
the production frame and since standard spin observables are defined most
naturally in the $\gamma N$ CM,  it is necessary to
 treat the decay in that same CM frame.
 In much of the literature the decay and density
matrix of the vector meson are described in the rest frame of
the vector meson and standard spin observables
are not invoked.
This difference in approach is partly a question of differing motivation.
Extraction of the vector meson's density
matrix in its rest frame, using either
the Gottfried-Jackson, Adair or Helicity  
axes, can provide a framework
for determining the character of
$t-$channel exchange mechanisms.
On the other hand,   efforts 
 to define the angular and energy evolution of
 standard spin observables, which could be driven
by specific resonances and other dynamical mechanisms, does require that
spin observables be defined in a 
consistent manner, especially for cases involving measuring
the vector meson decay along with having polarized photons
and/or nucleons (e.g. for double or triple spin observables). 
 Our purpose  is to define single and double
spin observables in a manner consistent with the treatment of the
decay process. Therefore, we describe all spin observables
and decays in the same overall $\gamma N$ CM frame.

  We begin by discussing the number and character 
of the basic
amplitudes needed to describe spin observables.
The photoproduction of vector mesons is described by twelve independent, complex helicity
amplitudes.
The  determination of these 12 amplitudes requires the measurement
 of twenty-three 
independent observables at each energy and angle. 
It is unlikely that such a complete set of vector meson photoproduction   
experiments will ever be accomplished. Nevertheless, it is natural that one should start by 
 measuring all single spin 
observables, plus the differential cross section. Let us consider the
number
of single spin observables, e.g. observables for which either the photon,
the target nucleon,  the recoil nucleon or the vector meson are polarized. 
With the $\hat{z}$ axis along the photon beam momentum, the $x$-$z$  plane
is the 
scattering plane, see Fig 1.  The vector meson is produced in the
final state $\hat{z}'$ direction,  with the normal
to the scattering plane remaining as $\hat{y}'=\hat{y},$
 see Fig. 1. 
For the photon,  the single spin observable is the photon asymmetry
 $\vec{\Sigma}^{\gamma},$    
(also called $\vec{A}^{\gamma}$),  
which has only a non-zero $\hat{x}$ component. 
The target nucleon single spin observable is called
the target polarization $\vec{T}^N$ 
  (also called the nucleon asymmetry $\vec{A}^N$~),  which has
only a non-zero $\hat{y}$ component. 
For the recoil nucleon there is the polarization $\vec{P}^{N'},$ 
which has only a non-zero $\hat{y}$ component. 
The vector meson has a vector polarization $\vec{P}^{V} = \hat{y} P^V_y$ and 
a tensor polarization $T^{V}_{ij}$ (in a cartesian basis).  
Thus, with the differential cross-section
$\sigma(\theta)$,
there are eight observables to consider 
before going on to double, triple, 
and even quadruple, spin observables,  which  
 require experiments with simultaneous 
information about two, three, and four particle spins. 
Experiments where one uses polarized photon beams and measures the vector
meson decays, fall into this category of double spin observables. 
In experiments where one has an unpolarized photon beam, an unpolarized 
target, and also one does not measure the recoil nucleon's polarization, 
the angular distribution of the two-body decay of 
the vector meson provides information about 
the single spin observables of the vector meson. 
For example, one could measure the angular distribution of the 
two decay pions for the case of $\rho$ meson photoproduction, or measure the 
two decay kaons for $\phi$ photoproduction.  
Such experiments are planned using the CLAS detector at TJNAF. 
What information about the vector meson is obtained from
 measuring the two-body decays?

In this paper, we first discuss the vector meson's single 
spin observables, obtainable with an unpolarized photon beam.   
Then we consider the case of a polarized photon beam
along with measurement of the vector meson's two-body decay modes; 
for this case one deals not with a 
single spin observable, but a double spin observable
or a spin correlation.  The full 
set of spin observables with  beam, target and/or recoil particles  
polarized was addressed in Ref.~\cite{Pich}.  Note that all spin observables
and the associated amplitudes and density matrices are defined
in the $\gamma N$ CM frame.

The eight independent single spin observables
for this reaction 
are: the differential cross-section $\sigma(\theta),$
the target asymmetry $ T^{N}_y,$
the photon beam asymmetry $  \Sigma^{\gamma}_x,$  
the recoil polarization  $P^{N'}_y,$
the vector meson's vector polarization $ P^{V}_y,$ 
and the vector meson's tensor polarization in a spherical tensor basis 
$T^{V}_{20}$, $T^{V}_{21}$,\ and \ $T^{V}_{22}$.   
Because of parity conservation, one has 
$P^{V}_{x'}  = 0$,  $P^{V}_{z'}  = 0$, 
$T^{V}_{2-2}  =  T^{V}_{22}$,  
and $T^{V}_{2-1}  = - T^{V}_{21}.$   
In  the language of density matrices,
measuring the four single spin observables of the vector meson
($P^{V}_y$, $T^{V}_{20}$, $T^{V}_{2 1}$, and $T^{V}_{2 2})$  
is equivalent to knowing all nine 
elements of the hermitean $3 \times 3$  spin density 
matrix $< 1\, \lambda  |\rho^V | 1\, \lambda' >,$
for the case of an unpolarized photon beam. 
To be specific these are the elements of $\rho^0_{\lambda \lambda '}$, 
defined in Ref.~\cite{Sch}, 
which formalism is often used in analysis of experimental data. 
 
With the CM amplitude for $\gamma + N \rightarrow V + N'$  denoted by 
${\bf T}$, 
the final CM density matrix is 
$\rho_f= {\bf T} \rho_i {\bf T}^\dagger = \frac{1}{4} {\bf T T}^\dagger,$
 where the $\frac{1}{4}$ arises from averaging over the initial unpolarized
photon and unpolarized target nucleon.
This $\rho_{\it{f}}$ reduces to $\frac{1}{2} \rho^V,$ where the final 
baryon's spin is to be summed over since it is not measured.
 The final density matrix then involves just the
spin state of the vector meson and is described by (see Ref.~\cite{Pich}):
\begin{equation}
\rho^{V} = \frac{1}{3}[ I + \frac{3}{2} \vec{S} \cdot \vec{P}^V 
+   \tau \cdot T^{V}] .
\label{rhopich}
\end{equation}  Note the above differs from the
definition used in Pichowsky et al.~\cite{Pich}, 
since we now require that in the limit of a pure state,
the polarization be limited to be at most one in magnitude
and the tensor polarization $T^V_{20}$ varies between
a maximum of $\frac{1}{\sqrt{2}}$
and a minimum of $-\sqrt{2}.$~\footnote{
Other changes from Ref.~\cite{Pich} are:
 $\vec{P}^V = {\rm Tr}( T T^\dagger \vec{S})/{\rm Tr} ( T T^\dagger),$
 $ T^V_{2 \mu}={\rm Tr}
 ( T T^\dagger \tau_{2 \mu})/{\rm Tr} ( T T^\dagger),$
and $\tau_{2 \mu}= \sqrt{3} [S_1 \times S_1]_{2 \mu}.$
These changes were made to  comply 
 with the Madison conventions. }
Since $\rho^V$ is a hermitian $3 \times 3$ matrix, Eq.(\ref{rhopich}) 
is a general form, 
where $\vec{S}$ is the spin-1 operator and $\tau$ is the symmetric 
traceless rank-2 operator with cartesian components 
$\tau_{ij} = \frac{3}{2}(S_i S_j + S_j S_i) - 2 \delta_{ij}$. 
On  a spherical basis
 $\tau \cdot T^{V}$ is defined as 
$\sum_{\mu} (-1)^{\mu}\, \tau_{2 -\mu}\ T^{V}_{2 \mu} 
 = \frac{1}{3} \sum_{i,j} \tau_{ij} T_{ij}.$ 
The function $\vec{P}^V$ defines the vector meson's vector
 polarization and $T^{V}_{2 \mu}$ the 
tensor polarization 
 of the vector meson,
 when both $T^{V}$ and the density matrix
are defined in the CM frame. 
We note that other authors discuss the elements of the 
frame dependent density matrix  
in the rest frame of the vector meson 
(see for example Refs.~\cite{Sch,Shim} and also Ref.~\cite{Gott} for $K^*$ 
production and decay).    
In our paper,  all single spin observables 
$\vec{P}^V$ and $T^V_{2 \mu}$ 
as well as double, triple, and quadruple spin 
correlations are defined only in the $\gamma N$ CM 
frame;  we do not invoke rest frames for individual
particles. The use of the overall $\gamma N$ CM frame also allows 
to exploit algebraic relations between various 
sets of spin observables which can be derived for vector meson
photoproduction. Such rules can be obtained as discussed in 
Ref.~\cite{Wen} and are extremely important in order to define 
complete sets of independent observables.

All of the spin observables depend on the squared total invariant 
energy $s$
and the $\gamma-V$ squared momentum transfer $t$ or, equivalently,
on the incident photon energy and the CM
scattering angles $\Theta, \Phi$
for the outgoing vector meson.~\footnote{We take
$\Phi\equiv 0$ for the scattering plane and
introduce azimuthal angles for the two-meson decay plane,
$\phi_1, \phi_2= \pi + \phi_1.$} 
That dependence is suppressed in using the notation
$T^V_{2\mu}$ instead of the full expression
$T^V_{2\mu}(E_\gamma, \Theta, \Phi),$  but should not be forgotten.
Rules for the angular and energy dependence were discussed in 
Ref.~\cite{Pich}
and could be used to analyze important underlying dynamics,
such as resonances, cusps, or missing state effects. 
The density matrix in the vector meson's rest frame
has been defined earlier by others using either the   
Gottfried-Jackson, Adair, or Helicity axes as shown in Fig. 2 
($\rho^{GJ}, \rho^A, \rho^H$). 
Extraction from data yields different values for 
$\rho^{GJ}, \rho^A, \rho^H$ (see for example Ref.\cite{Aach}), which 
are related by simple rotation matrices. 
These different  vector meson rest frame axes are useful for the study of exchange mechanisms 
and for study of helicity conservation, 
but are not useful for the definition of spin observables,
which requires consistent use of the $\gamma N$ CM system. 

\section{SINGLE SPIN OBSERVABLES FOR THE VECTOR MESON} 

To understand the connection
between the single spin observables of the vector meson 
and the density matrix elements in the $\gamma N$ CM system,
we express the matrix elements of the density matrix $\rho^{V}$ 
in terms of the meson's vector and tensor polarization by:
\begin{eqnarray}
\rho^{V} =& \frac{1}{3} 
\left(\begin{array}{ccc}

1 +  \frac{3}{2} P^{V}_{z'}+ \sqrt{\frac{1}{2}} T^{V}_{20}& 
\frac{3}{2} P^{V}_{1-1} +
 \sqrt{\frac{3}{2}} T^{V}_{2-1}&
\sqrt{3} T^{V}_{2-2}   \\
-\frac{3}{2}P^{V}_{11} -
 \sqrt{\frac{3}{2}} T^{V}_{21}& 
1- \sqrt{2} T^{V}_{20}   & 
\frac{3}{2} P^{V}_{1-1} 
-\sqrt{\frac{3}{2}} T^{V}_{2-1} \\
\sqrt{3} T^{V}_{22}&
-\frac{3}{2} P^{V}_{11} +
\sqrt{\frac{3}{2}} T^{V}_{21}  &
1- \frac{3}{2} P^{V}_{z'}+
 \sqrt{\frac{1}{2}} T^{V}_{20} 
       \end{array} \right)  .
\label{rho-elts}
\end{eqnarray} Here,$ 
P^{V}_{1\mu=\pm1}= \mp \sqrt{\frac{1}{2}}(P^{V}_{x'.}\pm i P^{V}_y).$ 
Because of the hermiticity of $\rho^{V}$ and parity conservation in the
reaction, the 
matrix elements in Eq.(\ref{rho-elts}) satisfy the conditions 
\begin{equation}
\rho^{V}_{\lambda \lambda '} = \rho^{V *}_{\lambda ' \lambda},
\ \ \ \ \ {\rm and }\ \ \ \ \ 
\rho^{V}_{\lambda \lambda '} = (-1)^{\lambda - \lambda '} 
\rho^{V}_{-\lambda -\lambda '},
\label{Hermiticity} 
\end{equation} 
where $\lambda, \lambda'$ are helicity labels and take the values 
+1, 0, -1. Therefore,  
$P^{V}_{x'}$ = $P^{V}_{z'}$ = 0, $T^{V}_{20}$, $T^{V}_{21}$, 
$T^{V}_{22}$ are all real, and   
$T^{V}_{2-1}  = -  T^{V}_{21}$, $T^{V}_{2-2} =
T^{V}_{22}$. 
The density matrix simplifies to 
\begin{eqnarray}
\rho^{V} =& \frac{1}{3} 
\left(\begin{array}{ccc}

1 + \sqrt{\frac{1}{2}} T^{V}_{20}& 
\frac{3}{2}\sqrt{\frac{1}{2}}(-i P^{V}_y) -
 \sqrt{\frac{3}{2}} T^{V}_{21}&
\sqrt{3} T^{V}_{22}   \\
\frac{3}{2}\sqrt{\frac{1}{2}}(i P^{V}_y)-
 \sqrt{\frac{3}{2}} T^{V}_{21}& 
1- \sqrt{2} T^{V}_{20}   & 
\frac{3}{2}\sqrt{\frac{1}{2}}(-i P^{V}_y)
+\sqrt{\frac{3}{2}} T^{V}_{21} \\
\sqrt{3} T^{V}_{22}&
\frac{3}{2}\sqrt{\frac{1}{2}}(i P^{V}_y)+
\sqrt{\frac{3}{2}} T^{V}_{21}  &
1+
 \sqrt{\frac{1}{2}} T^{V}_{20} 
       \end{array} \right)  .
\label{rhopc-elts}
\end{eqnarray} 
Note that  Tr $\rho^V=1.$ 
The $3 \times 3$ hermitian spin density matrix 
$\rho^{V}$ is described fully by four independent real quantities, 
$\rho^{V}_{00},$  $\rho^{V}_{1-1},$  Re($\rho^{V}_{10} $), 
and Im($\rho^{V}_{10}$). 
These four quantities in the $\gamma N$ CM frame are simply related to the
four vector-meson double spin observables 
$P^{V}_y$, $T^{V}_{20}$, $T^{V}_{21}$,  and  $T^{V}_{22}$ by:
\begin{eqnarray}
P^V_y &=&   -2 \sqrt{2}\ {\rm Im}\  \rho^V_{1 0}\nonumber  \\
T^V_{2 0}&=&\frac{1}{\sqrt{2}}\, (1  - 3\, \rho^V_{0 0} )\nonumber  \\
T^V_{2 1}&=& -  \sqrt{6}\ {\rm Re}\  \rho^V_{ 1 0}\nonumber \\
T^V_{2 2}&=& \sqrt{3}\  \rho^V_{ 1 -1} \  .  
\label{observrules} 
\end{eqnarray}

In  sections IV and V, we  discuss which of these four quantities can be determined
by measuring the angular distribution of the two mesons or leptons that arise from
 vector meson decay. 

\section{DOUBLE SPIN OBSERVABLES FOR THE VECTOR MESON WITH POLARIZED PHOTON BEAM} 

We now consider those double spin observables involving a polarized photon
beam and observation of the vector meson decay.  These are part of
a broader class of double spin observables with two particles
polarized; namely: 
beam-target $C^{\gamma N}$, 
beam-recoil $C^{\gamma N'}$, 
target-recoil $C^{N N'}$,
target-vector meson $C^{N V}$, 
and recoil-vector meson observables $C^{N' V}$.  
Here, we restrict ourselves to the 
beam-vector meson double spin observables $C^{\gamma V}$.
 
The final density matrix elements describing the vector meson 
is now related to the initial density matrix for
 the incident photon beam by:
$\rho_{f} = \frac{1}{2} {\bf T} \rho^\gamma {\bf T}^\dagger,$  where
the photon is described by
\begin{equation}
\rho^\gamma = \frac{I}{2}[ 1 + \vec{P}^S\cdot \vec{\sigma}^\gamma].
\label{rhogamma}
\end{equation}
Here, $\vec{P}^S$ is the Stokes ``vector," as discussed
 in Ref.~\cite{FTS}.
The final density matrix can be written as:
$\rho_{f} =\rho^V + P^S_x \rho^x + P^S_y \rho^y + P^S_z \rho^z.$
 The vector meson density matrices 
$\rho^{x}, \rho^{y},$ and $\rho^{z},$  which arise from the
 $\sigma^\gamma_x, \sigma^\gamma_y,$ and $\sigma^\gamma_z $ 
terms in $\rho^\gamma $ in Eq. (\ref{rhogamma}), 
are expressed in the $\gamma N$ CM frame in terms of standard 
beam-vector meson double
spin observables as:~\footnote{We have already invoked the
constraints due to parity and the hermiticity of the density matrix.}
\begin{equation}
\rho^x = \frac{1}{3} [ \Sigma_x^\gamma +
  \frac{3}{2}\, \sum_j S_j\  C^{\gamma V}_{x j} 
+ \  \sum_{\mu}\, (-1)^\mu \tau_{2 -\mu}\    C^{\gamma V}_{x 2 \mu}\, ] ,
\label{rhox-elts}
\end{equation} 
which parallels Eq. (\ref{rhopich}).

We now consider the case of $P_y^S = P_z^S = 0, P_x^S = \pm 1,$
which corresponds to a  photon linearly polarized
 perpendicular $(\hat{y})$ for $P_x^S = +1,$ or 
  parallel $(\hat{x})$ for $P_x^S = -1$ to the scattering plane.
When the decay products of the final vector meson are also measured,
 we have a final state
 density matrix $\rho^x.$    
Because of the hermiticity of $\rho^{x}$ and parity conservation, the 
matrix elements of $\rho^{x}$  satisfy the conditions 
\begin{equation}
\rho^{x}_{\lambda \lambda '} = \rho^{x *}_{\lambda ' \lambda},
\ \ \ \ \ {\rm and }\ \ \ \ \  
\rho^{x}_{\lambda \lambda '} = (-1)^{\lambda - \lambda '} 
\rho^{x}_{-\lambda -\lambda '}.
\label{xHermiticity} 
\end{equation} 

As a consequence, one finds that  
$C^{\gamma V}_{xx'}$ = $C^{\gamma V}_{xz'}$ = 0, $C^{\gamma V}_{x20}$, 
$C^{\gamma V}_{x21}$, 
$C^{\gamma V}_{x22}$ 
are all real, and   
$C^{\gamma V}_{x2-1}  = -  C^{\gamma V}_{x21}$, $C^{\gamma V}_{x2-2} =
C^{\gamma V}_{x22}$. 
The density matrix $\rho^{x}$ simplifies to 
\begin{eqnarray}
\rho^{x} =& \frac{1}{3}
\left(\begin{array}{ccc}

 \Sigma^{\gamma}_x + \sqrt{\frac{1}{2}} C^{\gamma V}_{x20}& 
 \frac{3}{2 \sqrt{2}} (-i C^{\gamma V}_{xy}) -  \sqrt{\frac{3}{2}} 
C^{\gamma V}_{x21}&
 \sqrt{3} C^{\gamma V}_{x22}   \\
\frac{3}{2 \sqrt{2}} ( i C^{\gamma V}_{xy}) -  \sqrt{\frac{3}{2}} 
C^{\gamma V}_{x21}& 
\Sigma^{\gamma}_x -   \sqrt{2}\, C^{\gamma V}_{x20}   & 
\frac{3}{2 \sqrt{2}} (- i C^{\gamma V}_{xy}) +  \sqrt{\frac{3}{2}}
C^{\gamma V}_{x21} \\
\sqrt{3} C^{\gamma V}_{x22}&
\frac{3}{2 \sqrt{2}}  ( i C^{\gamma V}_{xy}) +  \sqrt{\frac{3}{2}}
C^{\gamma V}_{x21}  &
\Sigma^{\gamma}_x + \sqrt{\frac{1}{2}} C^{\gamma V}_{x20} 
       \end{array} \right)  .
\label{rhoxpc-elts}
\end{eqnarray} 
Note that Tr$[\rho^x] =$ Tr$[ {\bf T} \frac{1}{2} \sigma^\gamma_x 
{\bf T}^\dagger ] 
= \Sigma^\gamma_x .$
This $3 \times 3$ hermitian spin density matrix 
$\rho^{x}$ is described fully by five 
independent real quantities, 
$\rho^{x}_{11}, \rho^{x}_{00},$  $\rho^{x}_{1-1},$  Re($\rho^{x}_{10} $), 
and Im($\rho^{x}_{10}$). 
These five quantities are related in the $\gamma N$ CM frame to the
four vector-meson double spin observables 
$C^{\gamma V}_{xy}$, $C^{\gamma V}_{x20}$, $C^{\gamma V}_{x21}$, 
 and 
$C^{\gamma V}_{x22}$
and the single spin observable $\Sigma^\gamma_x$ ~\footnote{
 The polarized photon asymmetry, 
$\Sigma^\gamma_x,$ is a single spin observable,
which enters in Eq. (\ref{rhoxpc-elts}) via the $P_x^S$ term of 
Eq. (\ref{rhogamma}.) } 
by:
\begin{eqnarray}
C^{\gamma V}_{x y}  &=&   -2 \sqrt{2}\, {\rm Im}\  \rho^x_{1 0} \nonumber \\
C^{\gamma V}_{x 2 0}&=&\frac{1}{\sqrt{2}}\,
 (\Sigma^\gamma_x  - 3\rho^x_{0 0} ) \nonumber \\
C^{\gamma V}_{x 2 1}&=& -  \sqrt{6}\ {\rm Re}\  \rho^x_{ 1 0} \nonumber \\
C^{\gamma V}_{x 2 2}&=&\sqrt{3} \rho^x_{ 1 -1}\  \nonumber  \\
\Sigma^\gamma_x &=& \rho_{00}^x + 2 \rho_{11}^x \  .  
\label{xobservrules} 
\end{eqnarray}

Next we consider the case of $P_x^S = P_z^S = 0, P_y^S = \pm 1,$
which corresponds to a photon linearly  polarized
at an angle of $\mp 45^\circ$ with respect to
the $\hat{x}$ axis. 
When the decay products of the final vector meson are also measured,  
the final state density matrix $\rho^y$ is:
\begin{equation}
\rho^y = \frac{1}{3} [ \Sigma_y^\gamma +
  \frac{3}{2}\, \sum_j S_j\  C^{\gamma V}_{y j} 
+ \  \sum_{\mu}\, (-1)^\mu \tau_{2 -\mu}\  
  C^{\gamma V}_{y 2 \mu}\, ] ,
\label{rhoy-elts}
\end{equation}
Because of the hermiticity of $\rho^{y}$ and parity conservation, the 
matrix elements in Eq.(\ref{rhoy-elts}) satisfy the conditions 
\begin{equation}
\rho^{y}_{\lambda \lambda '} =  \rho^{y *}_{\lambda ' \lambda},
\ \ \ \ \ {\rm and }\ \ \ \ \ 
\rho^{y}_{\lambda \lambda '} = - (-1)^{\lambda - \lambda '} 
\rho^{y}_{-\lambda -\lambda '}. 
\label{yHermiticity} 
\end{equation} 
Note that condition (3.7) for $\rho^y$ has the opposite sign from the
similar condition (3.3) for $\rho^x.$
As a consequence different observables are now constrained. 
One obtains, 
$\Sigma^{\gamma}_y$ = 0, 
$C^{\gamma V}_{yy} $ = 0, 
$C^{\gamma V}_{y20}$ = 0, 
$C^{\gamma V}_{yx'z'}$ = 0, 
$C^{\gamma V}_{yx'x'} - C^{\gamma V}_{yyy}$ = 0, 
$\sqrt{3}\ C^{\gamma V}_{y21} = - i C^{\gamma V}_{yyz'}$ , 
and   
$\sqrt{3}\ C^{\gamma V}_{y22} =   i C^{\gamma V}_{yx'y}$. 
Now $C^{\gamma V}_{y 21}$ and $C^{\gamma V}_{y 22}$ 
are both purely imaginary, and
$C^{\gamma V}_{y 2 -1}=  C^{\gamma V}_{y 2 1},$
$C^{\gamma V}_{y 2 -2}= -C^{\gamma V}_{y 2 2}. $

The density matrix simplifies to 
\begin{eqnarray}
\rho^{y} =& \frac{1}{3}
\left(\begin{array}{ccc}

\frac{3}{2}  C^{\gamma V}_{yz'}& 
\frac{3}{2 \sqrt{2}}   C^{\gamma V}_{yx'} +   \sqrt{\frac{3}{2}}
C^{\gamma V}_{y21}&
-\sqrt{3} C^{\gamma V}_{y22}   \\
\frac{3}{2 \sqrt{2}} C^{\gamma V}_{yx'} -  \sqrt{\frac{3}{2}} 
C^{\gamma V}_{y21}& 
0   & 
\frac{3}{2 \sqrt{2}} C^{\gamma V}_{yx'} -  \sqrt{\frac{3}{2}}
C^{\gamma V}_{y21} \\
\sqrt{3} C^{\gamma V}_{y22}&
\frac{3}{2 \sqrt{2}} C^{\gamma V}_{yx'} +  \sqrt{\frac{3}{2}}
C^{\gamma V}_{y21}  &
- \frac{3}{2} C^{\gamma V}_{yz'} 
       \end{array} \right)  .
\label{rhoypc-elts}
\end{eqnarray} 
Note that Tr$[\rho^y]$ = 0, because $\Sigma^{\gamma}_y$ = 0. 
 The $3 \times 3$ hermitian  density matrix 
$\rho^{y}$ is fully described by four independent real quantities, 
$\rho^{y}_{11},$  {\rm Im}\ ($\rho^{y}_{1-1}$),  Re($\rho^{y}_{10} $), 
and Im($\rho^{y}_{10}$). 
These four quantities are related in the $\gamma N$ CM frame to the
four vector-meson double spin observables 
$C^{\gamma V}_{yx'}$, $C^{\gamma V}_{yz'}$, $C^{\gamma V}_{y21}$, 
 and  $C^{\gamma V}_{y22}$ by:
\begin{eqnarray}
C^{\gamma V}_{y x'}  &=& 2\sqrt{2}\,   {\rm Re}\ (\rho^y_{1 0}) \nonumber \\
C^{\gamma V}_{y z'}  &=&  2\,      \rho^y_{1 1}  \nonumber \\
C^{\gamma V}_{y 2 1}&=&  \sqrt{6}\ i \ {\rm Im}\ (\rho^y_{1 0}) \nonumber \\
C^{\gamma V}_{y 2 2}&=&              - \sqrt{3} \rho^y_{1 -1}  \ .
\label{yobservrules} 
\end{eqnarray}

Finally,  we consider the case of $P_x^S = P_y^S = 0, P_z^S = \pm 1,$
which corresponds to a circularly polarized  photon,
with helicity $\pm 1,$ e.g. right or left circular polarization.   
With measurement of the decay products of the final vector meson, 
the final state density matrix $\rho^z$ is given by:

\begin{eqnarray}
\rho^{z} =& \frac{1}{3}
\left(\begin{array}{ccc}

\frac{3}{2}  C^{\gamma V}_{zz'}& 
\frac{3}{2 \sqrt{2}}   C^{\gamma V}_{zx'} +   \sqrt{\frac{3}{2}}
C^{\gamma V}_{z21}&
-\sqrt{3} C^{\gamma V}_{z22}   \\
\frac{3}{2 \sqrt{2}} C^{\gamma V}_{zx'} -  \sqrt{\frac{3}{2}} 
C^{\gamma V}_{z21}& 
0   & 
\frac{3}{2 \sqrt{2}} C^{\gamma V}_{zx'} -  \sqrt{\frac{3}{2}}
C^{\gamma V}_{z21} \\
\sqrt{3} C^{\gamma V}_{z22}&
\frac{3}{2 \sqrt{2}} C^{\gamma V}_{zx'} +  \sqrt{\frac{3}{2}}
C^{\gamma V}_{z21}  &
- \frac{3}{2} C^{\gamma V}_{zz'} 
       \end{array} \right)  .
\label{rhozpc-elts}
\end{eqnarray} 
 
Similarly,  
$\Sigma^{\gamma}_z$ = 0, 
$C^{\gamma V}_{zy} $ = 0, 
$C^{\gamma V}_{z20}$ = 0, 
$C^{\gamma V}_{zx'z'}$ = 0, 
$C^{\gamma V}_{zx'x'} - C^{\gamma V}_{zyy}$ = 0, 
$\sqrt{3}\ C^{\gamma V}_{z21} = - i C^{\gamma V}_{zyz'}$ , and   
$\sqrt{3}\ C^{\gamma V}_{z22} =   i C^{\gamma V}_{zx'y}$. 
Again $C^{\gamma V}_{z21}$ and $C^{\gamma V}_{z22}$ are purely imaginary 
and $C^{\gamma V}_{z2-1} = C^{\gamma V}_{z21}, C^{\gamma V}_{z2-2} = 
- C^{\gamma V}_{z22}$. 
The density matrix simplifies to 
\begin{eqnarray}
\rho^{z} =& \frac{1}{3}
\left(\begin{array}{ccc}

\frac{3}{2}  C^{\gamma V}_{zz'}& 
\frac{3}{2 \sqrt{2}}   C^{\gamma V}_{zx'} +   \sqrt{\frac{3}{2}}
C^{\gamma V}_{z21}&
-\sqrt{3} C^{\gamma V}_{z22}   \\
\frac{3}{2 \sqrt{2}} C^{\gamma V}_{zx'} -  \sqrt{\frac{3}{2}} 
C^{\gamma V}_{z21}& 
0   & 
\frac{3}{2 \sqrt{2}} C^{\gamma V}_{zx'} -  \sqrt{\frac{3}{2}}
C^{\gamma V}_{z21} \\
\sqrt{3} C^{\gamma V}_{z22}&
\frac{3}{2 \sqrt{2}} C^{\gamma V}_{zx'} +  \sqrt{\frac{3}{2}}
C^{\gamma V}_{z21}  &
- \frac{3}{2} C^{\gamma V}_{zz'} 
       \end{array} \right)  .
\label{rhozpc-elts}
\end{eqnarray} 
 The $3 \times 3$ hermitian spin density matrix 
$\rho^{z}$ is fully described by four independent real quantities, 
$\rho^{z}_{11},$  Im($\rho^{z}_{1-1}$),  Re($\rho^{z}_{10} $), 
and Im($\rho^{z}_{10}$). 
These four quantities are related in the $\gamma N$ CM frame to the
four vector-meson double spin observables 
$C^{\gamma V}_{zx'}$, $C^{\gamma V}_{zz'}$, $C^{\gamma V}_{z21}$, 
and $C^{\gamma V}_{z22}$ by:
\begin{eqnarray}
C^{\gamma V}_{z x'}  &=& 2\sqrt{2}\,   {\rm Re}\ (\rho^z_{1 0}) \nonumber \\
C^{\gamma V}_{z z'}  &=&  2\,      \rho^z_{1 1}  \nonumber \\
C^{\gamma V}_{z 2 1}&=&  \sqrt{6}\ i \ {\rm Im}\ (\rho^z_{1 0}) \nonumber \\
C^{\gamma V}_{z 2 2}&=&              - \sqrt{3} \rho^z_{1 -1}  \ .
\label{yobservrules} 
\end{eqnarray}

Which single and double spin observables can be measured
using the decay pions or leptons?

\section{ Pion Pair in Final State}

The $\rho$ meson is observed  via its decay products. 
For $\rho$ decay into a pion pair, one 
determines the angular distribution of the two pions in the
reaction 

\begin{equation}
\gamma + N    \rightarrow    \pi + \pi + N' ,  
\end{equation}
for example in the overall $\gamma N$ (or $\rho N'$) CM system (see Fig. 1). 
Here, we do not
deal with the nontrivial separation of the direct meson decay mode from
other mechanisms for producing the two
final pions.  To define spin observables that involve the $\rho$ meson,
such a separation is required. 
 In Appendix A,  we outline
how such background terms would enter into our discussion and the simple form
of the decay that results when one is allowed to ignore mechanisms
for two-meson production other than via the vector meson
production.   

 In this section, we
present the $\rho \rightarrow \pi \pi $ decay, but the discussion applies
also to the
case of $\phi \rightarrow K \bar{K}.$

\subsection{Decay Distributions for Single Spin Observables}

Assuming that the pions are produced solely via vector meson decay,
the two pion angular distribution 
in the overall $\gamma N$ (or $V N')$ CM frame 
is to be measured.   As shown in Appendix A, the two pion angular distribution
in the $V N'$ CM frame can be described using the angles
$\bar{\theta}$ and $\bar{\phi}$ between the direction of the
vector meson's momentum  $\vec{q}= q \hat{z}'$ and the vector $\vec{v}_1 - \vec{v}_2,$ where
$\vec{v}_1 = \vec{p}_1/E_1$ and $\vec{v}_2 = \vec{p}_2/E_2$
are the velocity vectors of the two decay mesons (e.g. $\pi_1$ 
and $\pi_2$).  
 The angles $\bar{\theta}, \bar{\phi}$ reduce, in the vector meson rest 
frame, to the angles of the decay pion $\pi_1$,  
e.g. $\bar{\theta}, \bar{\phi} \rightarrow \theta_1 , \phi_1.$       
(We note that the reference frame for these angles is 
$\hat{x}', \hat{y}, \hat{z}'$, 
where the $\hat{z}'$-axis is along the direction
 of the vector meson V.) 

For an unpolarized photon beam, the two-pion angular distribution is given by 

\begin{equation}
 dN^V = 
W^V(\bar{\theta}, \bar{\phi}) \ d \cos \bar{\theta}\  d \bar{\phi}. 
\end{equation} One counts the number of mesons at   $\bar{\theta}, \bar{\phi},$
 which entails measuring 
the momentum and energy of each meson, forming the two-meson
four vector $p_1 + p_2 = q$, and only select those pion pairs that
satisfy $q^2 = m_\rho^2$, allowing for the width of the
 vector meson.~\footnote{Boosting  
along the $\rho$ direction from the 
$\rho N'$ overall CM frame to the $\rho$ rest frame, one can show that
  the angle $\bar{\theta}$ is related to the $\rho$ rest frame  
angle $\theta_1$ of the decay pion $\pi_1$  by 
$\tan \bar{\theta} = \frac{E_{\rho}}{m_{\rho}} \tan \theta_1,$ 
while the azimuthal orientation of the scattering plane with respect
to the $\hat{z}'$ axis remains the same,  $\bar{\phi} = \phi_1$.} 

The decay distribution $W(\bar{\theta}, \bar{\phi})$ can be expressed in terms of the 
matrix elements of the $\gamma N$ CM density matrix $\rho^{V}$, and the CM decay 
amplitude $M$ (see Appendix A). 
The relation is $W^V(\bar{\theta},\, \bar{\phi}) = 
{\rm Tr}[M \rho_f M^\dagger ]$, 
where the trace appears because of summation over the vector 
meson helicities.   
Since we evaluate $\rho_f = {\bf T} \rho_i {\bf T}^{\dagger}$ in the 
$\gamma N$ CM frame, 
we need to evaluate the $\rho$ decay amplitude $M$ in the same overall CM 
frame (see Appendix A). Therefore,  we are dealing with the traditional 
frame in which spin observables are defined. 

Before parity conservation is imposed, one has~\cite{Sch}
 \begin{eqnarray}
W^V(\bar{\theta},\bar{\phi})&=&\frac{1}{4 \pi} \xi_V(\bar{\theta}) 
[
(1 - \sqrt{\frac{1}{2}}\, T^V_{20} (3 \cos^2 \bar{\theta} - 1) \nonumber \\
&+& \sqrt{3}\, {\rm Re}\  (T^V_{21}) \sin 2\bar{\theta} \cos \bar{\phi} + 
  \sqrt{3}\,  {\rm Im}\  (T^V_{21}) \sin 2\bar{\theta} \sin \bar{\phi} \nonumber \\
&-&\sqrt{3}\,{\rm Re}\ (T^V_{22}) \sin^2 \bar{\theta} \cos 2\bar{\phi}  
 -\sqrt{3}\, {\rm Im}\ (T^V_{22}) \sin^2 \bar{\theta} \sin 2\bar{\phi} ]. 
\label{Wpi} 
\end{eqnarray} The function  
$\xi_V(\bar{\theta})$ is a kinematical factor that arises from considering the
vector meson decay in the $\gamma N$ CM system.  It is expressed
in terms of the decay angle $\bar{\theta}$ as
\begin{equation}
\xi_V(\bar{\theta}) = \frac{1}{ 
(\sin^2\bar{\theta} + (\frac{E_\rho}{m_\rho})^2 
\cos^2\bar{\theta})^{5/2} }\  ,
\end{equation} where $m_\rho$ is the vector meson's mass
and $E_\rho$ is the vector meson's energy, which depends
on the production angles $\Theta, \Phi.$ 
For $ E_\rho \approx  m_\rho$ clearly $\xi_V(\bar{\theta}) \rightarrow 1,$ 
and the above expressions 
reduce to those obtained when the rest frame of the vector meson is
used along with the amplitude in the $\gamma N$ CM frame.
Therefore,  $\xi_V(\bar{\theta})$ is the major effect of a
consistent treatment of the decay and production,
along with the use of the decay angle $\bar{\theta}.$
We normalize the above angular distribution by dividing by the known 
factor $\xi_V$ and setting 
 $$\int d\bar{\Omega}\  W^V(\bar{\theta},\bar{\phi})
/ \xi_V(\bar{\theta}) \equiv 1$$.

Already at this stage the angular dependence $W^V(\bar{\theta} , \bar{\phi})$ 
does not depend on the vector polarization $P^{V}_y.$  Neither 
does the function $W^V$ depend on $P^{V}_{x'}$ nor $P^{V}_{z'}.$  
After parity conservation has been imposed, one has

\begin{eqnarray}
W^V(\bar{\theta},\bar{\phi}) &=& \frac{1}{4 \pi} \xi_V(\bar{\theta})\,  
\, 
[ 1 -   \sqrt{2}\  
T^V_{2\mu}(\Theta,\Phi)\  {\bf C}^*_{2\mu}(\bar{\theta}\bar{\phi})  ] \nonumber \\
&=& \frac{1}{4 \pi} \xi_V(\bar{\theta}) 
\, [ 1 -  \sqrt{\frac{1}{2}} T^V_{20} (3 \cos^2 \bar{\theta} - 1) + 
\sqrt{3}\ T^V_{2 1}\, \sin 2\bar{\theta} \cos \bar{\phi} - 
 \sqrt{3}\, T^V_{2 2}\, \sin^2 \bar{\theta} \cos 2\bar{\phi} ].
\label{WpiPC2} 
\end{eqnarray} Here ${\bf C}^*_{2\mu} \equiv 
\sqrt{\frac{4 \pi}{5} }\, Y^*_{2\mu} $ is a spherical
harmonic function. 
A general proof that $W^V(\bar{\theta},\bar{\phi})$ for
decay to two pseudoscalar mesons 
is independent of the vector meson's polarization $\vec{P}^V$
is presented in Appendix B. 

Both $P^{V}_{x'}$ and $P^{V}_{z'}$  vanish because of  parity conservation, 
but $P^{V}_y $ is non-zero, and contains all the information about the 
vector polarization of the vector meson. 
As seen from Eq. (\ref{WpiPC2}), the observable $P^{V}_y$ remains 
unmeasurable from the two meson decay 
mode.  

\subsection{Decay Distributions for Double Spin Observables}

For initial photon polarization,
the angular distribution of the decay mesons becomes
\begin{equation}
W(\bar{\theta} , \bar{\phi}) = W^V(\bar{\theta} , \bar{\phi}) 
+ P^S_x W^x(\bar{\theta} , \bar{\phi}) + P^S_y W^y(\bar{\theta} , \bar{\phi})
 + P^S_z W^z(\bar{\theta} , \bar{\phi}),
 \end{equation}
where $W^i(\bar{\theta} , \bar{\phi})= {\rm Tr} [ M \rho^i M^\dagger].$ 
Again both $\rho$ and $M$ are defined in the $\gamma N$ CM system. 
The explicit forms of $W^i(\bar{\theta}, \bar{\phi})$ are  
\begin{eqnarray}
W^x(\bar{\theta},\bar{\phi}) &=& \frac{1}{4 \pi} \xi_V(\bar{\theta})
 [\   \Sigma^{\gamma}_x -     
\sqrt{\frac{1}{2}}\, C^{\gamma V}_{x20} (3 \cos^2 \bar{\theta} - 1) + 
\sqrt{3}\  C^{\gamma V}_{x 2 1}\, \sin 2\bar{\theta} \cos \bar{\phi} - 
 \sqrt{3}\, C^{\gamma V}_{x 2 2}\, \sin^2 \bar{\theta} \cos 2\bar{\phi} \  ], \nonumber \\
 W^y(\bar{\theta},\bar{\phi}) &=& \frac{1}{4 \pi} \xi_V(\bar{\theta}) [\   
\sqrt{3}\ {\rm Im}(C^{\gamma V}_{y 2 1})\, \sin 2\bar{\theta} \sin \bar{\phi} - 
 \sqrt{3}\,{\rm Im}(C^{\gamma V}_{y 2 2})\, \sin^2 \bar{\theta} \sin 2\bar{\phi} \  ],
\nonumber \\
W^z(\bar{\theta},\bar{\phi}) &=& \frac{1}{4 \pi} \xi_V(\bar{\theta}) [\   
\sqrt{3}\  {\rm Im}(C^{\gamma V}_{z 2 1})\, \sin 2\bar{\theta} \sin \bar{\phi} - 
 \sqrt{3}\,{\rm Im}(C^{\gamma V}_{z 2 2})\, \sin^2 \bar{\theta} \sin 2\bar{\phi} \  ].
\label{WzpiPC2} 
\end{eqnarray} 
Here, $\Sigma^{\gamma}_x$ is a single spin observable (the polarized photon
asymmetry); all other observables in Eq. (\ref{WzpiPC2}),  
$C^{\gamma V}_{x20}, C^{\gamma V}_{x 2 1}, 
C^{\gamma V}_{x 2 2}, 
C^{\gamma V}_{y 2 1},C^{\gamma V}_{y 2 2},
C^{\gamma V}_{z 2 1},C^{\gamma V}_{z 2 2},$ involve polarized photons
and the tensor polarization
of the vector meson and therefore 
 are double spin observables.

Again one concludes that the double
spin observables
$C^{\gamma V}_{xy}, 
C^{\gamma V}_{yx'}, C^{\gamma V}_{yz'}, 
C^{\gamma V}_{zx'}, C^{\gamma V}_{zz'},$ 
which involve a polarized photon and 
the vector polarization of the vector meson, 
do not contribute and  
cannot be measured by 
meson decay distributions.

\section{ Lepton Pair in Final State}

The $\rho$ meson can also decay into a pair of leptons. 
For $\rho$ decay into an electron-positron pair one has the reaction 
\begin{equation}
\gamma + N    \rightarrow    e^+ + e^- + N',  
\end{equation} 
for which one might determine the angular distribution of the two decay
 leptons in the $\gamma N$ CM system. 
The vector meson's decay into  a muon pair 
\begin{equation}
\gamma + N    \rightarrow    \mu^+ + \mu^- + N'  ,
\end{equation}
has a similar angular distribution.
The angular distribution of the two leptons can be expressed in  terms of 
the matrix elements of the vector meson's density matrix $\rho^V.$ 

\subsection{Decay Distributions for Single Spin Observables} 

The
angular distribution of the decay lepton pair for the case that the lepton
spins are not measured and the photon
is not polarized is given by (see Appendix B):
 
\begin{eqnarray}
W^V(\bar{\theta},\bar{\phi}) &=& \frac{1}{8 \pi} 
\, [ 3\ \xi_V^{3/5}(\bar{\theta})
 +\xi_V(\bar{\theta})\{-1 +  \sqrt{2} 
T^V_{2\mu}(\Theta,\Phi) {\bf C}^*_{2\mu}(\bar{\theta}\bar{\phi})\}] \nonumber \\
 &=& \frac{1}{8 \pi} 
\, [ 3\, \xi_V^{3/5}(\bar{\theta}) +\xi_V(\bar{\theta})\{-1 + 
\sqrt{\frac{1}{2}}\, T^V_{2 0} (3 \cos^2 \bar{\theta} - 1)\nonumber \\ 
 &-&   \sqrt{3}\ T^V_{2 1}\ \sin 2 \bar{\theta} \cos   \bar{\phi} 
 +     \sqrt{3}\ T^V_{2 2}\ \sin^2 \bar{\theta} \cos 2 \bar{\phi}\}].
\label{IpiPC2} 
\end{eqnarray} 
Note that the angular dependence $W^V(\bar{\theta} , \bar{\phi})$  
is  independent of the  vector meson's vector
 polarization  $\vec{P}^{V}, $  
 but does depend on the 
vector meson's tensor polarization $T^V_{2 \mu}$.  Proof of this assertion
is given in Appendix B. 

This should be contrasted to the fact that complete spin information 
(vector polarization as well as tensor polarization) 
about $W$ and $Z$ vector bosons can 
be extracted from their leptonic decay angular distributions. For example, 
Ref.~\cite{Shim} deals with vector meson weak decay 
of the type $W^- \rightarrow \ell^- \bar{\nu}_\ell$ 
or $Z \rightarrow \ell^- \ell^+$ in which the
parity violation term $ 1 \pm \gamma_5$ yields S, P  and D-wave interference
terms that depend on the vector meson's vector polarization.
Therefore, angular distributions of weak decays of vector mesons, 
with the lepton spin unmeasured, do depend on the vector polarization.
The difference with the processes discussed in this paper is that 
in vector meson photoproduction 
the $\rho$ and $\phi$ meson decays are strong or electro-magnetic 
interactions that conserve parity.  
However, one should note that the operator $\frac{1}{2}(1 \pm \gamma_5)$ 
also serves as a spin projection operator. 
This indicates that if we can measure the spin of
one of the decay leptons, 
in the parity conserving $\rho$ or $\phi$ leptonic decay, 
the dependence on the vector polarization will appear also.  
Indeed,  that is the result  for the
case of two-lepton decay with one lepton spin measured.
 
\subsubsection{Final Leptonic Spin}

If one can measure the spin of one of the final leptons,
 the helicity projection operator
$ \frac{1}{2}(1 \pm \gamma_5)$ is introduced,
 where for example the + sign means 
projecting out a right-handed electron or a left-handed positron. 
 The angular distribution of the
decay leptons now does depend on the
vector meson's vector polarization $P^V_y :$

\begin{equation}
\tilde{W}^V(\bar{\theta},\bar{\phi}) =
W^V(\bar{\theta}, \bar{\phi}) \mp \xi_V(\bar{\theta}) 
\frac{1}{4 \pi} P^V_y\ \sin \bar{\theta} \sin \bar{\phi} ,  
\label{IspiPC2} 
\end{equation} 
where the $\mp$ sign corresponds to the $\pm$ sign in 
the spin projection operator. Here, $W^V(\bar{\theta}, \bar{\phi})$ is 
given in Eq. (\ref{IpiPC2}) for the case where no lepton spin is measured. 

 The above results are for the case of no photon polarization and therefore
involve single spin observables.  Now 
consider the case of  beam-vector meson double spin observables.
  
\subsection{Decay Distributions for Double Spin Observables}
For initial photon polarization,
the angular distribution of the decay leptons, without measuring their 
spin, becomes

\begin{equation}
W(\bar{\theta} , \bar{\phi}) = W^V(\bar{\theta} , \bar{\phi}) 
+ P^S_x W^x(\bar{\theta} , \bar{\phi}) 
+ P^S_y W^y(\bar{\theta} , \bar{\phi}) 
+ P^S_z W^z(\bar{\theta} , \bar{\phi}),
\label{Wlepton}
\end{equation}
where $W^i(\bar{\theta} , \bar{\phi})= {\rm Tr} [ M \rho^i M^\dagger].$
 The explicit forms of $W^i(\bar{\theta} ,\bar{\phi})$ are  

\begin{eqnarray}
W^x(\bar{\theta},\bar{\phi}) &=& \frac{1}{8 \pi} 
[3\ \xi_V^{3/5}(\bar{\theta})\, \Sigma^{\gamma}_x +
 \xi_V(\bar{\theta}) \{-\Sigma^{\gamma}_x +
 \sqrt{\frac{1}{2}} \, C^{\gamma V}_{x 2 0} 
(3\cos^2\bar{\theta} - 1) \nonumber \\
 &-& \sqrt{3}\ C^{\gamma V}_{x 2 1}\ \sin 2 \bar{\theta} \cos \bar{\phi}    
 + \sqrt{3}\ C^{\gamma V}_{x 2 2}\ \sin^2 \bar{\theta} \cos 2 \bar{\phi}\}], \nonumber \\
W^y(\bar{\theta},\bar{\phi}) &=& \frac{1}{8 \pi}  \xi_V(\bar{\theta})[ 
 - \sqrt{3}\ {\rm Im}(C^{\gamma V}_{y 2 1})\ \sin 2\bar{\theta} \sin \bar{\phi}    
 + \sqrt{3}\ {\rm Im}(C^{\gamma V}_{y 2 2})\ \sin^2 \bar{\theta} \sin 2 \bar{\phi}], \nonumber \\
W^z(\bar{\theta},\bar{\phi}) &=& \frac{1}{8 \pi}  \xi_V(\bar{\theta})[ 
 - \sqrt{3}\ {\rm Im}(C^{\gamma V}_{z 2 1})\ \sin 2\bar{\theta} \sin \bar{\phi}    
 + \sqrt{3}\ {\rm Im}(C^{\gamma V}_{z 2 2})\ \sin^2 \bar{\theta} \sin 2 \bar{\phi}]. 
\label{IypiPC2} 
\end{eqnarray}  
Here, $\Sigma^{\gamma}_x$ is a single spin observable (the polarized photon
asymmetry); the 7 other observables in Eq. (\ref{Wlepton}),  
$C^{\gamma V}_{x20}, C^{\gamma V}_{x 2 1}, C^{\gamma V}_{x 2 2}, 
C^{\gamma V}_{y 2 1},C^{\gamma V}_{y 2 2},
C^{\gamma V}_{z 2 1},C^{\gamma V}_{z 2 2},$ 
involve polarized photons and tensor polarization
of the vector meson and are double spin observables.

\subsubsection{Final Leptonic Spin}

The 5 double spin observables
$C^{\gamma V}_{xy}, 
C^{\gamma V}_{yx'}, C^{\gamma V}_{yz'}, 
C^{\gamma V}_{zx'}, C^{\gamma V}_{zz'},$ 
which involve the vector polarization of the vector meson, 
do not contribute to the lepton decay distribution (see 
Eq. (\ref{IypiPC2}))  unless one measures the
spin of one of the decay leptons or has a weak interaction.

If a lepton spin is determined, all $W^i(\bar{\theta}, \bar{\phi})$ in 
Eq. (\ref{Wlepton}) are replaced by $\tilde{W}^i(\bar{\theta}, \bar{\phi})$, 
where 
\begin{eqnarray}
\tilde{W}^x(\bar{\theta},\bar{\phi}) &=& W^x(\bar{\theta},\bar{\phi})  
\mp  \xi_V(\bar{\theta}) 
\frac{1}{ 4 \pi} C^{\gamma V}_{xy}\ \sin \bar{\theta} \sin \bar{\phi}  
 ,  \nonumber \\
\tilde{W}^y(\bar{\theta},\bar{\phi}) &=& W^y(\bar{\theta},\bar{\phi}) 
 \mp \xi_V(\bar{\theta}) 
\frac{1}{ 4 \pi} \, [ C^{\gamma V}_{y z'} \cos \bar{\theta}   
 +  C^{\gamma V}_{yx'}\ \sin \bar{\theta} \cos \bar{\phi} ], \nonumber \\
\tilde{W}^z(\bar{\theta},\bar{\phi}) &=& W^z(\bar{\theta},\bar{\phi}) 
 \mp \xi_V(\bar{\theta}) 
\frac{1}{ 4 \pi} \, [ C^{\gamma V}_{z z'} \cos \bar{\theta}   
 +  C^{\gamma V}_{zx'}\ \sin \bar{\theta} \cos \bar{\phi} ].  \nonumber \\
\label{IyspiPC2} 
\end{eqnarray} 
Therefore,  from the angular distribution one can now extract double
spin observables that involve the vector meson's vector polarization
 $C^{\gamma V}_{xy}, 
C^{\gamma V}_{yx'}, C^{\gamma V}_{yz'}, 
C^{\gamma V}_{zx'}, C^{\gamma V}_{zz'}.$ 
However, the measurement of two-lepton decays,
which is already difficult by virtue of the 
$10^{-4} - 10^{-5}$ decay branching
ratio  to the dominant two meson mode,  is
made essentially impossible by the
need to measure the spin of a decay lepton.
Hence, for practical purposes the vector meson's vector polarization remains
unmeasurable, even for the case of two lepton decay.
 
The general discussion of why the vector polarization does not enter into the
$W^V(\bar{\theta},\bar{\phi})$ angular distribution, 
when no final lepton spin is measured,
is given in Appendix B.

\section{ Conclusion}

In pionic and leptonic parity-conserving two-body decays
of vector mesons,  we have shown that measurement of the final decay angular
distributions yields only the tensor polarization of the vector meson.
Access to the vector meson's vector polarization
is possible if one measures the final state leptonic spin (or
if one has access to a weak decay),  but both are presently not feasible.
Hence,  one of the simpler single spin observables,
$P^V_y$ is difficult to
measure and we must learn to live without it. 
The basic dynamics must be extracted from other observables. 
One possibility for finding $P_y^V$ is
to ascertain an interference mechanism with the background
that could isolate vector polarization effects.  Another
possibility, albeit remote, is to measure the final leptonic spin.
Finally,  using the algebra of measurement contained in 
Ref.~\cite{Pich,Wen} and
associated Fierz transformations,  it might be
possible to obtain the vector polarization from sets of
other single, double, triple, etc., spin observables,  which can be
measured.  That step requires a solution of the
problem of determining a complete set of measurements
for vector meson photoproduction.  

It has been shown in Appendices A and B that single spin observables involving
measuring the decay pions or leptons (without lepton spin measurements)
 do not yield
information about the vector meson's vector polarization,  but
only depend on the tensor polarization.
Vector mesons with two-body decays that conserve parity, are therefore
self-analyzing with respect to their tensor polarization.~\footnote{  
The procedures discussed in Appendices A and B also imply 
 that spin-2 mesons with parity conserving two-body decays can be  
self-analyzing with respect to their rank-2 and rank-4 tensor polarization,
but do not reveal their vector or octupole polarizations via the
angular distributions of their decay products.   }

The proof in Appendices A and B also shows that all double, triple,
and quadruple spin observables involving 
the vector polarization of the vector meson cannot be
obtained from processes where the final vector meson decays  into
two mesons.  Therefore, within the set of single
and double spin observables there are 16 spin observables that are
inaccessible, unless one measures the lepton spin in 
a leptonic decay mode.  These inaccessible observables are:
one single spin observable: 
$ P^V_y;$ 
five beam-vector meson double spin observables:
$ C^{\gamma V}_{xy}\ , 
C^{\gamma V}_{yx'}\ ,C^{\gamma V}_{yz'}\ ,
C^{\gamma V}_{zx'}\ , C^{\gamma V}_{zz'}\ ;$
five target-vector meson double spin observables: 
$C^{N V}_{xx'}\ , C^{N V}_{xz'}\ , 
C^{N V}_{yy}\ , 
C^{N V}_{zx'}\ , C^{N V}_{zz'}\ ;$ 
and five recoil nucleon-vector meson double spin observables:
$C^{N' V}_{xx'}\ , C^{N' V}_{xz'}\ , 
C^{N' V}_{yy}\ ,
C^{N' V}_{zx'}\ , C^{N' V}_{zz'}\ .$

Which observables are accessible with
present experimental techniques? 
For an unpolarized photon beam, 
just by measuring the angular 
distribution of the decay mesons or the decay leptons, one does find the 
values of the 3 single tensor polarization observables of the vector meson 
$(T_{20}, T_{21}, T_{22})$. 
For a polarized photon beam, one finds 7 double (photon-vector meson) 
spin observables 
$C^{\gamma V}_{x 20}\ ,  C^{\gamma V}_{x 21}\ ,  C^{\gamma V}_{x 22}\ ,  
 C^{\gamma V}_{y 21}\ ,  C^{\gamma V}_{y 22}\ ,  
C^{\gamma V}_{z 21}\ ,  C^{\gamma V}_{z 22}\ , $  
 involving the polarization of the photon
 and the tensor 
polarization (but again not the vector polarization) of the vector meson. 
For a polarized target, the decay angular distribution determines 7 double 
(target-vector meson) spin observables,
$C^{N V}_{x 21}\ ,  C^{N V}_{x 22}\ ,  
C^{N V}_{y 20}\ ,  C^{N V}_{y 21}\ ,  C^{N V}_{y 22}\ , 
C^{N V}_{z 21}\ ,  C^{N V}_{z 22}\ , $ 
and by detecting the recoil 
polarization and the angular decay distribution one can obtain 7 double 
(recoil-vector meson) spin observables  
$C^{N' V}_{x 21}\ ,  C^{N' V}_{x 22}\ ,  
C^{N' V}_{y 20}\ ,  C^{N' V}_{y 21}\ ,  C^{N' V}_{y 22}\ , 
C^{N' V}_{z 21}\ ,  C^{N' V}_{z 22}\ . $ 
Recall that all these observables are related to the spin density 
matrices in the $\gamma N$ CM frame, and are functions
of the incident beam energy and of the
scattering angles $\Theta, \Phi$ of the vector meson. 
There is therefore still a very rich source of information to be 
obtained from the angular distribution of the vector meson's decay 
products, if one does a careful analysis of the data. 

As part of that careful analysis,  if one wishes to
extract spin observables, then both the vector meson
production amplitude and its decay must be
expressed in the $\gamma N$ CM system.  As a consequence,
a kinematic factor $\xi_V(\bar{\theta})$ must be included.

\acknowledgments
The authors wish to thank
Professors S.A.Dytman, J. Mueller, R. Schumaker, P. Cole,
W. Roberts and M. Ripani for stimulating their
interest in this problem.
One author (W.M.K.) thanks the University of Pittsburgh,
another (F.T.) thanks Rutgers University for 
warm hospitality.  This research was 
supported, in part, by the U.S. National Science Foundation
Phy-9504866(Pitt) and Phy-9722088(Rutgers). 
One of the authors(W-T. C) has been supported by an Andrew Mellon 
Predoctoral Fellowship. 

\begin{appendix}
\section{Role of Background Terms}
In this appendix we outline the role of background terms,
such as arise from direct production of
two mesons without intermediate vector meson production.
By ignoring this background, we obtain the simplified version for the
decay angular distribution used in this paper and in Ref.~\cite{Sch}.

The amplitude for the photoproduction
of two pseudoscalar mesons has the general form
\begin{equation}
{\bf T}=  M {\bf T}^V    + {\bf T}^B,
\end{equation}  
where ${\bf T}^V$
describes the photoproduction of a vector meson and
$M$ describes the subsequent decay of the vector meson into
two pseudoscalar mesons.  The term ${\bf T}^B$ describes other ``background"
mechanisms for the direct production of, for example,
 two pions.~\cite{Rob}~\cite{Oset} 
 We assume that
it will be possible to select only those pion pairs with the
correct kinematics of having $\rho-$like attributes.
Namely, in the CM frame of the two pions, 
they should be in a pure relative P-wave 
(the quantum number of the $\rho$) and the
energy of both pions should satisfy 
$E_{pion1}=E_{pion2}= m_\rho/2$.  
Satisfying these two
constraints does not guarantee that only the $ M {\bf T}^V $ term
contributes;  one could get direct production contributions from
${\bf T}^B$ without an intermediate $\rho.$  In the $\rho$ rest frame the
decay amplitude $M_\lambda \propto 
 {\cal D}^1_{\lambda 0}(\theta_1, \phi_1)^* .$ 
Here $\theta_1, \phi_1$ are the angles of one of the decay pions 
in the $\rho$ rest frame 
(where the momentum of the recoil nucleon is
 along the $-\hat{z}'$ axis).

The two-body decay amplitude $M_\lambda$ of $\rho$ into $\pi_1$ given by 
$(E_1, \vec{p}_1)$ and $\pi_2$ given 
by $(E_2, \vec{p}_2)$ has the relativistic invariant form 

\begin{equation}
M_{\lambda} = - i g_{\rho \pi \pi} \varepsilon ^{\mu}(q,\lambda) 
(p_1 -p_2)_{\mu},
\label{Mform}
\end{equation} 
where $\varepsilon^{\mu}(q,\lambda)$ is the polarization vector of the 
$\rho$ meson. We can therefore express the decay
in the $\gamma N$ CM frame using Eq.(\ref{Mform}). 
Introducing the relative velocity of the decay products 
$\vec{v} = \vec{v}_1-\vec{v}_2$, one can write $M_{\lambda}$ 
as a 3-vector dot product and a helicity independent kinematic factor 

\begin{equation}
M_{\lambda} = i g_{\rho \pi \pi} \frac{2 E_1 E_2}{E_1+E_2} 
\vec{\varepsilon }(q,\lambda).(\vec{v}_1-\vec{v}_2)
\\
  = i g_{\rho \pi \pi} \frac{2 E_1 E_2}{E_1+E_2} 
{|\vec{v}_1-\vec{v}_2|} {\cal D}^1_{\lambda 0} (\bar{\theta}, \bar{\phi})^*.
\label{Mpion}
\end{equation} 
The above form of $M_\lambda$ holds in the $\rho$ rest frame and in the $\gamma N$ 
CM frame, and $\bar{\theta}$ and $\bar{\phi}$ are the angles 
(see Fig. 2) between $\hat{z}'$ and the relative velocity 
$\vec{v} = \vec{v}_1-\vec{v}_2$. 
In the $\rho$ rest frame Eq.(\ref{Mpion}) reduces to 
\begin{equation}
M_{\lambda} = i g_{\rho \pi \pi}\ 2 p_1\   
{\cal D}^1_{\lambda 0} (\theta_1, \phi_1)^*.
\end{equation} 

For an unpolarized beam,  an unpolarized target and with no
measurement made of the
recoil nucleon's polarization,  the reaction
$\gamma + N \rightarrow \pi_1 + \pi_2 + N'$ is
described by  the transition probability
\begin{equation}
{\bf T}{\bf T}^\dagger = (M {\bf T}^V + {\bf T}^B) (M {\bf T}^V + {\bf T}^B)^\dagger =
  M {\bf T}^{V} {\bf T}^{V \dagger} M^\dagger + 
  M {\bf T}^{V} {\bf T}^{B \dagger} +
 {\bf T}^{B} {\bf T}^{V \dagger} M^\dagger +
 {\bf T}^{B} {\bf T}^{B \dagger} .
\end{equation} 
In our discussion, we have assumed ${\bf T}^B=0.$  Using that assumption, 
we can describe the angular distribution of the decay pions in the 
overall CM frame by 
\begin{equation}
W^V(\bar{\theta},\bar{\phi}) = \rm{Tr}[ \it{M} \rho^V \it{M}^\dagger ],
\end{equation} 
where
\begin{equation}
\rho^V = {\bf T}^{V} {\bf T}^{V \dagger}.
\end{equation} 
Note that $M_\lambda$ is given in the $\gamma N$ CM frame by Eq. (A3).
The kinematic factor $\xi_V(\bar{\theta})$ in the text arises by combining:
\begin{eqnarray}
(\frac{2 E_1 E_2}{E_1+E_2}
{|\vec{v}_1-\vec{v}_2|} )^2  
&=&  \frac{4\, p_1^2}{ \sin^2 \bar{\theta}
 + (\frac{E_q}{m_\rho})^2 \cos^2 \bar{\theta}} ,  
\end{eqnarray} with the invariant density of state factors
$$\frac{d^3 p_1}{ 2 E_1} \  \frac{d^3 p_2}{ 2 E_2}
\  \delta^4(q-p_1-p_2).$$
 Here $\bar{\theta}$ is the
angle between the velocity difference vector
$\vec{v}_1-\vec{v}_2$ and the vector meson's momentum
$\vec{q} = q \hat{z}'$
in the $\gamma N$ CM system.

\section{Meson and Lepton Decay Angular Distribution}

Under the assumption that background terms can be neglected, 
 we now demonstrate that the vector polarization,  both for
pionic and leptonic decays,  does not appear in the decay
angular distribution. This has already been shown in the text.  
Here, we show that the result is due to
general symmetry properties.  

The expression (A3), yields for the
pionic decay case:  
\begin{equation}
W^V(\bar{\theta},\bar{\phi}) = C \xi_V(\bar{\theta}) 
  \sum_{\lambda, \lambda'}  
{\cal D}^1_{\lambda 0}(\bar{\theta}, \bar{\phi})  
{\cal D}^1_{\lambda' 0}(\bar{\theta}, \bar{\phi})
\times
(-1)^{\lambda'} 
< 1 \lambda | [ I + \frac{3}{2} \vec{S} \cdot \vec{P^{V}} + 
 \tau \cdot T^{V}]| 1 -\lambda' >, 
\end{equation}  
where $C $ incorporates the decay coupling and kinematic factors,  
which are independent of the helicities and the
angles $\bar{\theta}, \bar{\phi}.$
Note that ${\cal D}^1_{\lambda 0}(\bar{\theta}, \bar{\phi})  
{\cal D}^1_{\lambda'0}~(\bar{\theta}, \bar{\phi})$
is a symmetric function of the $\rho$ helicity indices
$\lambda,\lambda'$; 
whereas,
$(-1)^{\lambda'} < 1 \lambda | \vec{S}| 1 -\lambda' >$ 
is an antisymmetric function of  
$\lambda,\lambda',$  using the Wigner-Eckart theorem. 
Thus, the term involving the vector polarization $\vec{P}^{V}$ vanishes 
in the description of the purely two pion decay mode. 
If the background term ${\bf T}^B$ is re-introduced,
then there would be terms in $W(\bar{\theta},\bar{\phi})$ for which
$\vec{P}^{V}$ plays a role.  

The above conclusion also applies to the lepton pair decay when
the spins of the decay leptons are not measured.
The decay amplitude $M_{\lambda, \lambda_+,\lambda_-}$
for the $1^-$ decay of the leptons,
\footnote{Two leptons, $\mu^+ \mu^-$, 
in a $1^-$ state (the quantum numbers of the $\rho$),  must be in
either a triplet S- or triplet D-wave. A P-wave (singlet or triplet) 
would form positive parity states $0^+$, $1^+$, or $2^+$, 
and therefore is not allowed.}
using a $ \bar{\psi}\gamma^\mu \psi \rho_\mu$ coupling~\footnote{
The decay angular  distribution with addition of a
tensor coupling also does not depend on the
vector polarization of the vector meson.
} for lepton-pair to vector meson equals   
\begin{equation}
M_{\lambda, \lambda_+,\lambda_-}=
- i g_{\rho l_1 l_2} \varepsilon^{\mu}(q,\lambda) 
\bar{u}(p_-,\lambda_-) \gamma_{\mu} v(p_+,\lambda_+),
\end{equation}
where the constant $g_{\rho l_1 l_2}$
includes all factors and is fixed by the decay width of the $\rho$ 
meson into a lepton pair ($l^+_1 l^-_2$). The lepton's  helicities are 
$\lambda_+$ and $\lambda_-$,   
and $\lambda$ is the vector meson's helicity. 
In forming the decay distribution,
we need the combination 
 $M_{\lambda, \lambda_+,\lambda_-} M^*_{\lambda', \lambda_+,\lambda_-}$
 where $M$ is the decay amplitude in the overall CM frame.  

Summing over the unobserved final lepton helicities
and using the usual projection operator rules, the above reduces to a 
trace; the result is:

\begin{eqnarray}
\sum_{\lambda_+,\lambda_-}\ 
M_{\lambda, \lambda_+,\lambda_-} M^*_{\lambda', \lambda_+,\lambda_-}
&=& \frac{g^2}{4 m^2}
\varepsilon^{\mu}(q,\lambda) \varepsilon^{* \nu}(q,\lambda') 
{\rm Tr} [(\pslash_2 + m) \gamma_\mu (\pslash_1-m) \gamma_\nu ] \nonumber\\
&=& \frac{g^2 m^2_{\rho}}{2 m^2} \delta_{\lambda \lambda'} 
- \frac{2 g^2}{m^2} (\frac{E_1 E_2}{E_1 + E_2})^2 
\vec{\varepsilon}(q,\lambda).(\vec{v}_1 - \vec{v}_2)
\vec{\varepsilon}^{\ *} (q,\lambda').(\vec{v}_1 - \vec{v}_2) .
\end{eqnarray} In evaluating
Eq.(B3) terms of order$ (\frac{m}{m_\rho})^2$ are ignored. 
As for the meson decay  $\bar{\theta}$ and $\bar{\phi}$ are the angles between $\hat{z}'$ 
and the relative velocity 
of the lepton pair $\vec{v} = \vec{v}_1 - \vec{v}_2$. 
The angular distribution of the lepton pair is now described by:

\begin{equation}
W^V(\bar{\theta},\bar{\phi}) = \sum_{\lambda } 
\sum_{\lambda_+,\lambda_-}\ 
M_{\lambda, \lambda_+,\lambda_-} M^*_{\lambda',\lambda_+,\lambda_-}
< 1 -\lambda | [ I  + \frac{3}{2}~ 
\vec{S} \cdot \vec{P}^V +  \tau \cdot T^{V}]| 1 \lambda' >, 
\end{equation} 
where 
$\sum_{\lambda_+,\lambda_-}\ 
M_{\lambda, \lambda_+,\lambda_-} M^*_{\lambda',\lambda_+,\lambda_-} 
(-1)^{\lambda'}$ 
{\it again} is a symmetric function of $\lambda, \lambda'$. 
Thus the vector polarization also does not contribute to the
angular distribution of the leptons,  when
the lepton spins are summed over, in the absence of 
a background term.

If one of the leptonic spins is measured then the above reasoning does not apply
and terms involving the vector meson's
vector polarization appear.

\section{Spin-dependent Lepton Decay Angular Distribution}

We mention an interesting similarity in the description of the 
parity conserving decay $V \rightarrow \ell^+ \ell^-$ relevant in this 
paper 
and the parity non-conserving decays  $Z \rightarrow \ell^+ \ell^-$ or 
$W^- \rightarrow \ell^- \bar{\nu}_\ell$. 
This similarity demonstrates why in the weak decay the angular 
distribution alone gives complete information on the polarization 
of the vector meson, while in the electromagnetic decay one needs an 
additional measurement of the lepton helicity to obtain the same 
complete spin information. 

We start from a Lorentz covariant form of the density matrix  
$\rho^{\mu \nu}$ related to the previously discussed density matrix 
$\rho^{\lambda \lambda'}$ by

\begin{equation}
\rho^{\mu \nu} = \sum_{\lambda \lambda'}\, \varepsilon ^\mu(q,\lambda)\, 
\rho^{\lambda \lambda'}\, \varepsilon ^{*\ \nu}(q,\lambda'), 
\end{equation} 
or
\begin{equation}
\rho^{\lambda \lambda' } = \sum_{\mu \nu}\, 
\varepsilon ^*_\mu(q,\lambda)\, 
\rho^{\mu \nu}\, \varepsilon _\nu(q,\lambda'). 
\end{equation} 

If we define the decay amplitude into leptons with respective momenta 
$\vec{p}_1$ and $\vec{p}_2$ and spin projections $m_1$ and $m_2$ as 
\begin{equation}
M_{\lambda}^{m1,m2} = \sum_{\mu} \ L_{\mu}^{m1,m2}\,
 \varepsilon ^\mu(q,\lambda), 
\end{equation} 
where 
\begin{equation}
L_{\mu}^{m1,m2} = g \bar{u}_{m1}(p_1) \gamma_\mu (a -b \gamma_5)  
v_{m2}(p_2), 
\end{equation}where the parameters $a$ and $b$ will
be discussed below. 
The angular distribution for definite $m_1$ and $m_2$ is found from 
\begin{eqnarray}
W^{m1,m2}(\bar{\theta}, \bar{\phi}) &=&
\sum_{\lambda \lambda'}\ M_{\lambda}^{m1,m2}\ \rho^{\lambda \lambda' }\  
M^{\dagger \ m1,m2}_{\lambda'}  \nonumber \\
&=&
\sum_{\lambda \lambda'}\ \sum_{\mu \nu}\ 
L_{\mu}^{m1,m2}\,  \varepsilon ^\mu(q,\lambda)\,
 \rho^{\lambda \lambda' }\ \varepsilon ^{*\ \nu}(q,\lambda')\, 
L_{\nu}^{m1,m2 *}\,  \nonumber \\
&=&
 \sum_{\mu \nu}\ \rho^{\mu \nu}\,  
L_{\mu}^{m1,m2}\, L_{\nu}^{m1,m2 *}\,  .
\label{Wm1m2}
\end{eqnarray} We can now define the Lorentz tensor 
\begin{eqnarray}
L_{\mu \nu}^{m1,m2}\, 
  &\equiv& L_{\mu}^{m1,m2}\, L_{\nu}^{m1,m2 *}\, \nonumber \\
&=&  
g^2 \bar{u}_{m1}(p_1) \gamma_\mu (a -b \gamma_5)  v_{m2}(p_2)
  \bar{v}_{m2}(p_2) \gamma_\nu (a -b \gamma_5)  u_{m1}(p_1).\ 
\end{eqnarray} Summing over spin projections
 $m_1$ and $m_2$ leads to
$W(\bar{\theta},\bar{\phi})= \sum_{m1 m2}\,
 W^{\ m1,m2}(\bar{\theta},\bar{\phi})$ and 
 $L_{\mu \nu} = 
\sum_{m1 m2} L_{\mu \nu}^{m1,m2}$ and  

\begin{eqnarray}
L_{\mu \nu}    &=&  
g^2 {\rm Tr} [\bar{u}(p_1) \gamma_\mu (a -b \gamma_5)  v(p_2)
  \bar{v}(p_2) \gamma_\nu (a -b \gamma_5)  u(p_1)\ ]
\nonumber \\
 &=& 
\frac{ g^2}{4 m^2} {\rm Tr} [(\pslash_1 + m) \gamma_\mu 
(a -b \gamma_5) (\pslash_2-m) \gamma_\nu (a -b \gamma_5)  ].
\label{Lmunu}
\end{eqnarray}

The Lorentz tensor $L_{\mu \nu}$ defined in Eq. (\ref{Lmunu}) 
describes all three cases that are 
relevant for the discussion in this Appendix. 
For $a = 1$ and $b = 0$, $L_{\mu \nu}$ describes 
the angular distribution $W(\bar{\theta},\bar{\phi})$
 of parity-conserving leptonic decay 
(for example $\rho \rightarrow \ell^+ \ell^-$), 
where one sums over the lepton helicities. 
Because the spin projection operator has the form 
$ \frac{1}{2}( 1 \pm \gamma_5)$, 
the choice $a = \frac{1}{2}$, $b = \pm \frac{1}{2}$, 
 in $L_{\mu \nu}$ 
corresponds with parity conserving leptonic decay where $\ell^-$ is 
respectively purely right handed or  purely left handed. 
For $a = b = 1$, the same $L_{\mu \nu}$ describes 
the angular distribution 
of a parity non-conserving weak decay, where the lepton helicities 
are not observed 
(for example in a process where $Z \rightarrow \ell^+ \ell^-$). 
From the above expressions for $L_{\mu \nu}$ in Eq. (\ref{Lmunu}) and 
$W(\bar{\theta}, \bar{\phi})$  
of Eq. (\ref{Wm1m2}) one can obtain the angular distribution of the decay 
leptons in each of the three cases.
  Only in the last two cases the angular distribution
 $W(\bar{\theta}, \bar{\phi})$ contains information about
the vector polarization of the vector meson due to the presence in Eq.~(C7)
of terms linear in $\gamma_5.$ 

\end{appendix}


%

\input epsf
\begin{figure}
\epsfysize= 7in 
\epsfbox{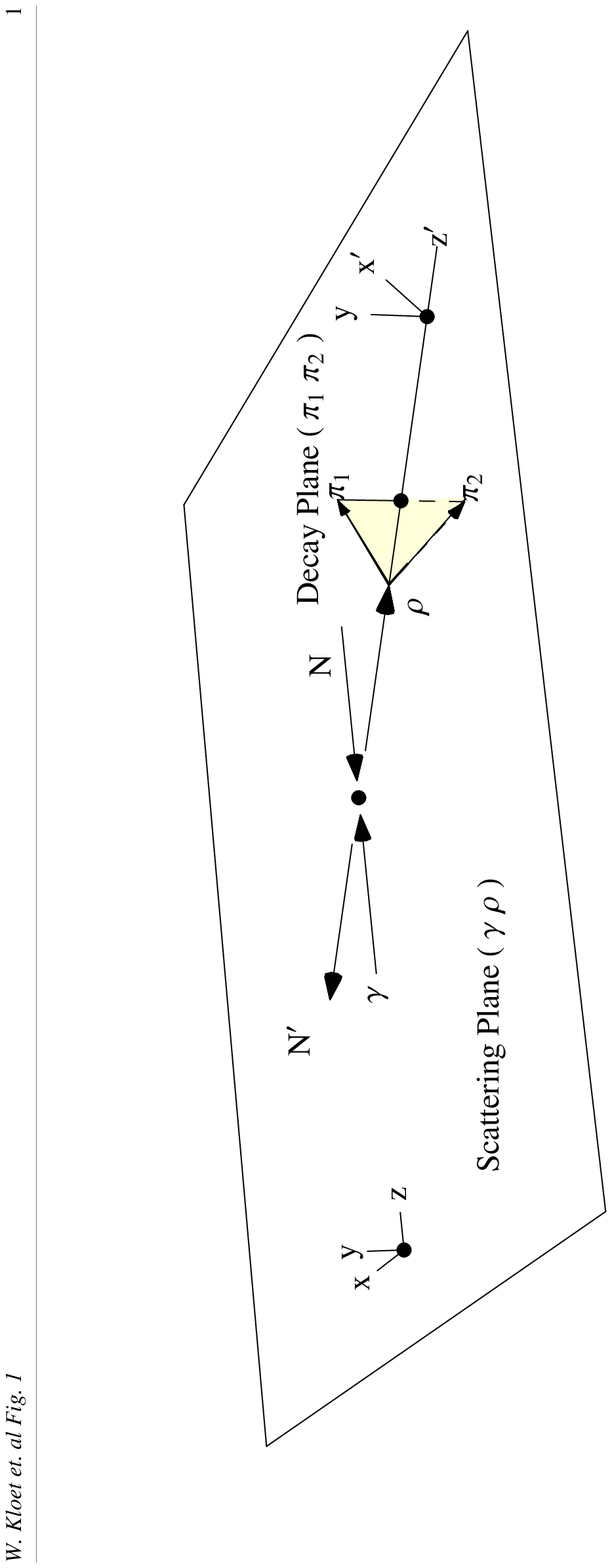}    
\caption{ Kinematics for the $\gamma N \rightarrow V  N'$ reaction,
where the incident photon is in the $\hat{z}$ direction,
the vector meson$(V=\rho)$ is produced in the $\hat{z}' $ direction
at the scattering angle $\Theta,\Phi$ and
the normal to the scattering plane is in the $\hat{y}$ direction.
The decay plane of either the two mesons $\pi_1,\pi_2$ or two leptons is
shown in the $\gamma N$ CM system,  where the 
decay products are in the
$\theta_1, \phi_1$ and $\theta_2, \phi_2$  
directions.}
\end{figure}
\newpage
\begin{figure}
\epsfxsize= 5in 
\epsfbox{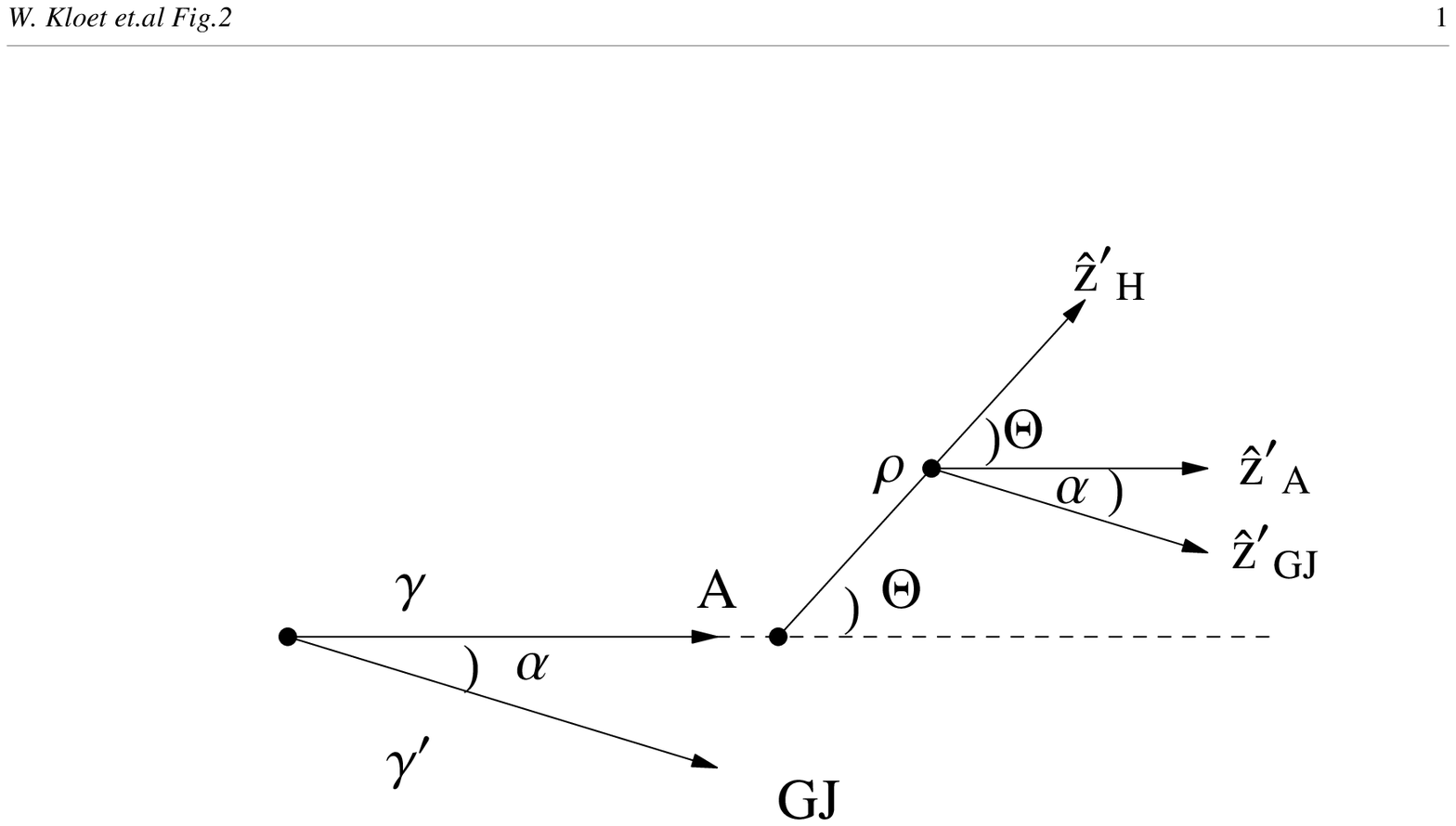}    
%
\caption{ The Gottfried-Jackson (GJ), Adair(A), and
Helicity(H) choice of axes are shown.  All three choices
are defined in the vector meson's rest frame and are related to
each other by simple rotations (by angles
$\alpha $ or $\Theta$) about their common 
 normal to the scattering plane, e.g. about $\hat{y}' .$
The H choice is defined by the
direction of the vector meson's momentum $\vec{q}$
and the normal $\hat{y}'$ to the scattering
plane (both of which remain the same for the
vector meson rest and the $\gamma N$ CM frames).
The GJ choice also uses the normal $\hat{y}'$ to the scattering
plane, but  the $\hat{z}'_{GJ}$ direction in the
vector meson's rest frame is defined by the incident photon's
direction--as seen from the vector meson's rest frame.  That last step
entails a Lorentz transformation so that the photon's direction
is seen along the vector $\gamma'$
 at the angle $\alpha$ as shown in the figure. 
The A choice again uses the common
normal to the plane, plus the original (untransformed)
 photon direction.  Each choice has its advantages,
but they are not useful for defining single and double 
spin observables,  which involve the photoproduction amplitude in 
the $\gamma N$ CM system.  
 }
\end{figure}
\newpage
\begin{figure}[t]
\epsfxsize= 5in 
\epsfbox{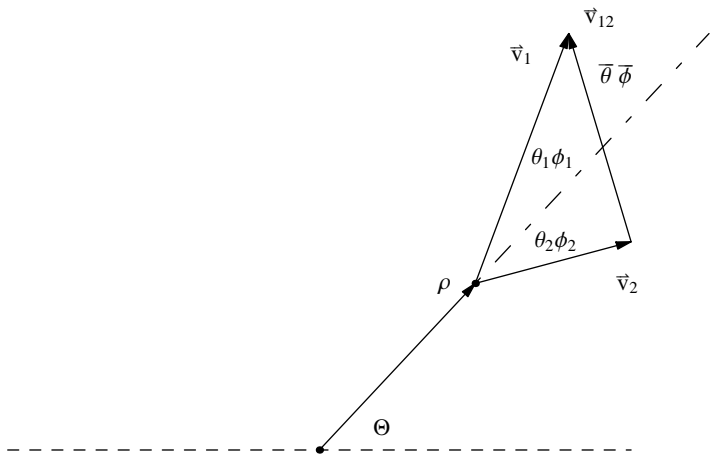}    

\caption{ A vector meson$(V=\rho)$ is produced in the $\hat{z}',$ direction
at the scattering angle $\Theta,\Phi,$  where we take the
scattering plane to have $ \Phi\equiv 0 .$  
The  velocities $\vec{v}_1= \vec{p}_1/E_1, \vec{v}_2= \vec{p}_2/E_2,$
  of the two mesons (or two leptons) are
shown in the $\gamma N$ CM system,  where they
are in the
$\theta_1, \phi_1$ and $\theta_2, \phi_2 = \pi + \phi_1$ directions.
 Note $\bar{\theta},
\bar{\phi}$ are defined as the angles between the
velocity difference vector
$\vec{v}_1 - \vec{v}_2$  and the direction of the
vector meson $\vec{q}= q \hat{z}' ,$ all in the $\gamma N$ CM system. 
The decay plane is out of the
scattering plane by an azimuthal angle $\bar{\phi},$ 
which equals $\phi_1.$ } 
\end{figure}

\end{document}